%% file: krause_etal.2009_aph.tex
\documentclass[useAMS,usenatbib,usegraphicx]{mn2e}

\usepackage{mathptmx}
\include{def}
\title[Cluster magnetic fields with the SKA]{Measurements of the Cosmological
       Evolution of Magnetic Fields with the Square Kilometre Array}

\author[M.\ Krause et al.]{Martin Krause,$^{1,2,3}$\thanks{E-mail:
           {\tt M.Krause@mrao.cam.ac.uk, krause@mpe.mpg.de, mkrause@usm.lmu.de}}
        Paul Alexander,$^{1,4}$ Rosie Bolton,$^1$ J\"orn Geisb\"usch,$^1$
        \newauthor David A.\ Green$^1$ and Julia Riley$^1$\\
        $^1$Astrophysics Group, Cavendish Laboratory, 19~J.~J.~Thomson Avenue,
            Cambridge CB3 0HE\\
        $^2$Max-Planck-Institut f\"ur Extraterrestrische Physik,
            Giessenbachstrasse, 85748 Garching, Germany \\
        $^3$Universit\"atssternwarte M\"unchen, Scheinerstr.~1,
            81679 M\"unchen, Germany\\
        $^4$Kavli Institute for Cosmology Cambridge, Madingley Road,
        Cambridge, CB3 0HA}
\begin{document}

\date{Accepted 2009 August 5.  Received 2009 July 6; 
  in original form 2009 April 3}

\pagerange{\pageref{firstpage}--\pageref{lastpage}} \pubyear{2009}

\maketitle

\label{firstpage}

\begin{abstract}
We investigate the potential of the Square Kilometre Array (SKA) for measuring
the magnetic fields in clusters of galaxies via Faraday rotation of background
polarised sources. The populations of clusters and radio sources are derived
from an analytical cosmological model, combined with an extrapolation of
current observational constraints. We adopt an empirical model for the Faraday
screen in individual clusters, gauged to observations of nearby clusters and
extrapolate the polarisation properties for the radio source population from
the NRAO VLA Sky Survey. We find that about 10~per~cent of the sky is covered
by a significant extragalactic Faraday screen. Most of it has rotation measures
between 10 and 100~rad~m$^{-2}$. We argue that the cluster centres should have
up to  about 5000~rad~m$^{-2}$. We show that the proposed mid frequency
aperture array of the SKA 
\change{as well as the lowest band of the SKA dish array}
are well suited to make measurements for most of these
rotation measure values, typically requiring a signal-to-noise of ten. We
calculate the spacing of sources forming a grid for the purpose of measuring
foreground rotation measures: it reaches a spacing of 36~arcsec for a 100~hour
SKA observation per field. We also calculate the statistics for background RM
measurements in clusters of galaxies. We find that a first phase of the SKA
would allow us to take stacking experiments out to high redshifts ($>1$), and
provide improved magnetic field structure measurements for individual nearby
clusters. The full SKA aperture array would be able to make very detailed
magnetic field structure measurements of clusters with more than 100 background
sources per cluster up to a redshift of 0.5 and more than 1000 background
sources per cluster for nearby clusters, and could for reasonable assumptions
about future measurements of electron densities in high redshift clusters
constrain the power law index for the magnetic field evolution to better than
${\Delta}\change{m}=0.4$, if the magnetic field in clusters should follow
$B\propto(1+z)^\change{m}$.
\end{abstract}

\begin{keywords}
surveys --- cosmology: observations --- galaxies: clusters: general ---
magnetic fields --- radio continuum: general --- instrumentation: polarimeters
\end{keywords}

\begin{table*}
\begin{minipage}{15.0cm}
\caption{Comparison of radio telescopes --- derived properties.}\label{telcomp}
\begin{tabular}{@{}p{4.5cm}@{\hspace{4mm}}rrrrr@{}}\hline
	Name		& VLA$^\mathrm{a}$\footnotetext{$^\mathrm{a}$~See:
{\tt http://www.vla.nrao.edu/astro/guides/vlas/current/}.}
		& SKA~AA 	& SKA~Dishes	& SKA~Dishes Phase~1 \\ \hline
%
%
\mbox{$A_\mathrm{eff}/T_\mathrm{sys}\,$}$^\mathrm{b}$\footnotetext{$^\mathrm{b}$~Effective
collecting area over system temperature.}
                        & 70--180        & 10,000       & 10,000
                        & 1,200 \\
	\begin{tabular}{@{}l@{}}frequencies of interest	/ GHz\end{tabular}
			& \begin{tabular}{@{}r@{}}0.3--0.34, 1.24--1.7,\\ 4.5--5, 8.1--8.8\end{tabular}
                                        & 0.3--1	        & 0.8--10	& 0.8--10        	\\
	instantaneous bandwidth / MHz	& 86		& 700		& 250		& 250			\\
	1-h sensitivity / $\mu$Jy
			& 17		& 0.18		& 0.29		& 2.5			\\
\begin{tabular}{@{}l@{}}field of view / square degree$^\mathrm{c}$\footnotetext{$^\mathrm{c}$~For
the dish telescopes, given at the bottom of the frequency range; $f$ denotes the observing frequency in GHz.}
\end{tabular}
			& $(0.7/f)^2$	& $250$ 	& $2$	        & $~20^\mathrm{d}$\footnotetext{$^\mathrm{d}$~Assuming phased array feeds.} \\ \hline
\end{tabular}
\end{minipage}
\end{table*}

\section{Introduction}

The plasma between the galaxies in clusters is magnetised. Observations at
radio frequencies are our main quantitative probe of cluster magnetic fields.
This is done by three main methods: equipartition arguments applied to cluster
halo radio sources \citep[e.g.][]{Gioea93}, comparison of X-ray Inverse
Compton (IC) with the radio synchrotron data in radio sources
\citep[e.g.][]{BPL98}, and Faraday rotation of polarised radio sources within
or behind the cluster, combined with X-ray data to determine the electron
densities \citep[][]{Kimea91,CKB01}. The results have been reviewed regularly
\citep{CT02,GF04,FG07}. Galaxy clusters are found to possess magnetic fields of
typically a few, up to of order 10~$\mu$G.

The structure within cluster magnetic fields has been deduced from Faraday
rotation of extended cluster radio sources
\citep{Murgea04,VE05,Govea06,Guiea08}. For the few cases observed so far, the
data is consistent with a Kolmogorov-type power spectrum. Because the field is
frozen into the plasma in the high conductivity limit, the structure of the
field directly traces the underlying kinematics of the gas.

Current studies of magnetic fields in clusters are limited to a few nearby
objects ($z < 0.1$). Hence, we have no means to answer questions regarding the
origin and cosmological evolution of these fields. Correspondingly, few
cosmological structure formation simulations have included magnetic fields. To
date, none have the resolution to address directly the field evolution in
galaxy clusters. However, from the eddy turnover times, and the high efficiency
of the turbulent dynamo, one might expect that the magnetic energy should
always be in equipartition with the turbulent kinetic energy (Ryu, private
communication). Hence, studying the cosmological evolution of fields in
clusters might not only lead to the origin of magnetism, but also trace cluster
kinematics.

Magnetic fields in clusters of galaxies might be connected to the cooling flow
problem. If cooling flows in nearby objects are quenched by weak jets, as has
been often suggested \citep[e.g.][]{KB03}, the magnetic fields will be
amplified up to equipartition with the enhanced kinetic energy level provided
by the very same jets.

\change{
The Faraday rotation method is expected to benefit substantially from
new developments of radio telescopes over the coming decade. The
Australian ASKAP project, as well as the South African MEERKAT project
are expected to start operation within a few years – both instruments
will have sensitivities competitive with the eVLA, (enhanced Very
Large Array) but with substantially higher survey speed at frequencies
below 1.4GHz.  The Square Kilometre Array (SKA) is the next generation
radio interferometer and will be a transformational instrument with an
improvement in sensitivity compared to the eVLA of a factor of fifty
with a $10^5$ increase in survey speed.  At frequencies below 500MHz,
aperture arrays will form the collector with dishes at frequencies
above 1.4 GHz.  In the mid frequency range (0.3 – 1.4 GHz) large
survey speeds will be achieved by using aperture arrays or dishes with
focal-plane arrays \citep{Schilea07}.  Phase 1 of the telescope
(with between 10 and 15 percent of the collecting area) is scheduled
for operation in about 2015 with phased deployment of the rest of the
telescope over the next few years and with completion in 2021/22.
Aperture arrays provide the most exciting technology for the
mid-frequency range with the largest increase in survey speed provided
they can meet an appropriate performance to cost \citep{AF09}.  
The derived properties for the SKA, based on \citet{Schilea07} and 
\citet{AF09} are summarised in Table~\ref{telcomp},
together with a comparison to the VLA.}

In this paper, we investigate the potential of these instruments, in particular
the SKA, to measure cosmological evolution in cluster magnetic fields. We
concentrate on Faraday rotation studies: the magnetic field along the line of
sight rotates the polarisation angle by
\eq{\Delta \chi = \frac{\mathrm{RM}}{(1+z)^2} \lambda^2,}
where $\lambda$ is the observing wavelength and we included the cosmological
factor ($z$ is the redshift of the rotating Faraday screen) to account for the
redshift of the radio emission traveling from the Faraday screen to us. This
factor ensures that for background sources with redshifts much higher than the
targeted cluster, the contribution to the rotation of the polarisation angle
from the vicinity of the radio source will be negligible (unless that RM
contribution is extremely high), since it occurs at much higher frequencies. RM
is related to the thermal electron density, $n_\mathrm{e}$ (cm$^{-3}$), and the
line-of-sight magnetic field, $B$ ($\mu$G), as:
\eq{\mathrm{RM} = 812 \int_0^L n_\mathrm{e}  {  B} \cdot
  \mathrm{d}l \; \mathrm{rad\, m^{-2}},}
where the path-length d$l$ is measured in kpc.

The rotation measure has contributions from all along the line of sight, from
the inter-galactic medium (IGM) in the vicinity of the source itself, from the
intra-cluster medium (ICM) in the clusters along the line of sight, and from
our own galaxy. The galactic contribution typically dominates at galactic
latitudes below about $|b|<20^\circ$. Above $|b|>40^\circ$, the galactic
contribution becomes smaller than $|\mathrm{RM}|<30 \,\mathrm{rad\, m^{-2}}$
\citep{SK80}, and can be as small as $5 \,\mathrm{rad\, m^{-2}}$
\citep{Guiea08}. Hence, high rotation measure from extragalactic sources at
high galactic latitudes can most likely be attributed to Faraday rotation
within the gas in galaxy clusters.

\subsection{Observations to date}

So far, the only individual cluster with a significant number of observed
polarised members {\em and} background sources is the Coma cluster. Seven
sources within 35~arcmin display a clearly enhanced $|\mathrm{RM}|$ between
35~and 65~rad~m$^{-2}$ \citep{Kimea90,Ferea95}. These sources, as well as
control sources in the vicinity of the cluster, but avoiding sightlines through
Coma's ICM, have been observed in many individual pointings with the VLA. For
the cluster A514, six embedded radio sources have been observed in individual
VLA pointings. These show a decline of $|\mathrm{RM}_\mathrm{max}|$ from
154~rad~m$^{-2}$ in the centre of the cluster to 15~rad~m$^{-2}$ on the
outskirts. Several other clusters, each with a few polarised sources, have been
observed with similar results \citep{Ferea99,Taylea01,Govea06,Guiea08}. Many
clusters have only one or two sufficiently bright radio sources in or behind
them\footnote{See: {\tt
http://www.mpa-garching.mpg.de/{\char'176}kdolag/BCluster/}.}. Therefore
stacking has been used to infer a general RM~value for nearby clusters
\citep{CKB01}. A systematic increase in $|\mathrm{RM}|$ is found towards the
cluster cores.

First attempts have been made to infer the magnetic field structure via the
power-spectrum with Faraday rotation studies. For the cluster A2255,
\citet{Govea06} measure the Faraday rotation of four embedded radio sources.
For different assumptions about the magnetic field power spectrum they can
extrapolate this data to predict the synchrotron emission of the radio halo.
Comparison to observations of the radio halo then constrains the power law
index of the magnetic field's 3D-power spectrum to the range three to four,
consistent with Kolmogorov turbulence, with a possible spatial variation. The
model also predicts polarisation of the halo, which is indeed detected.

This paper is structured as follows: First, we discuss the simulation method in
Section~\ref{sims}. The Faraday rotation detection statistics and the
implications for the measurement of the cosmological evolution of the magnetic
field in clusters are presented in Section~\ref{res}. We discuss our findings
in Section~\ref{disc}, and summarize them in Section~\ref{conc}.
%
%
Throughout, we use the following cosmological parameters: $h=0.7$, $\Omega_m=
0.3$, $\Omega_\Lambda=0.7$.

\begin{figure}
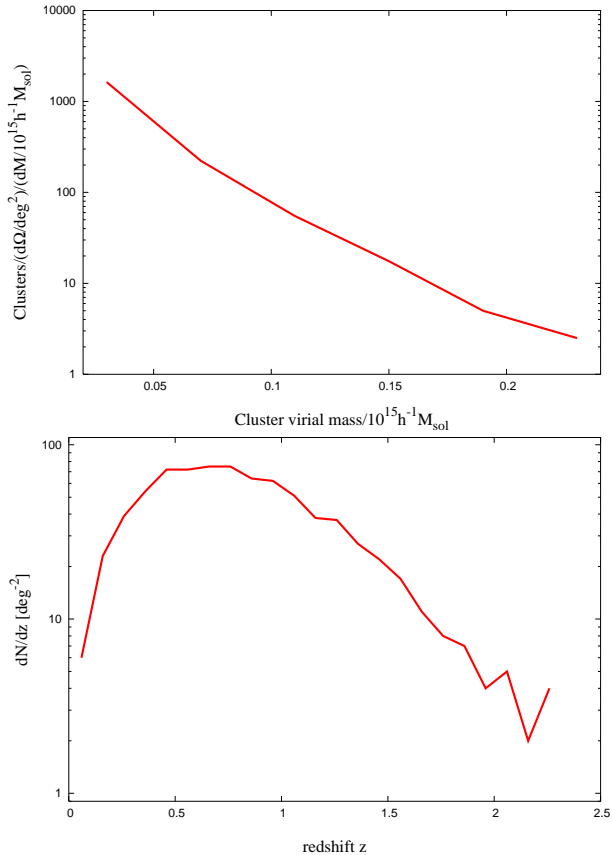

\centerline{\includegraphics[angle=270,width=8.0cm]{dNdM.epsi}}
\centerline{\includegraphics[angle=270,width=8.0cm]{dNdz.epsi}}
\caption{\small Mass (top) and redshift (bottom) distribution of the adopted
galaxy cluster model. See text for details of the model.}\label{cls_props}
\end{figure}

\begin{figure}
\centerline{\includegraphics[width=8.0cm]{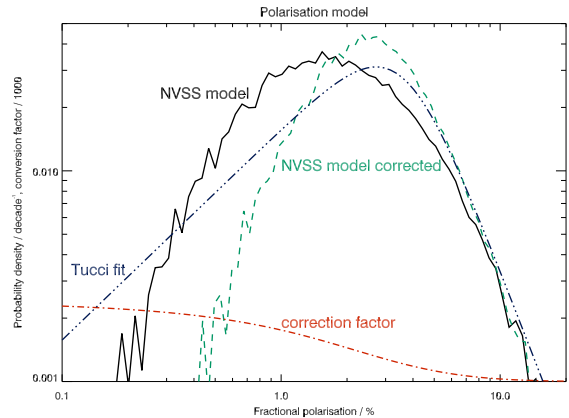}}
\caption{Distribution of the fractional polarisation for the radio source
model. The dark blue triple dot-dashed line represents the fit from
\citet{Tucea03} (not yet corrected for Faraday depolarisation). The solid black
line represents our adaptation of this fit. This model is corrected for Faraday
depolarisation by the polarisation dependent correction factor shown as red
dot-dashed curve (divided by 1000). The model finally adopted in this paper is
shown as a green dashed line. The polarisation distribution functions agree all
at high fractional polarisations, were they are gauged against NVSS
data.}\label{fpol}
\end{figure}

\begin{figure*}
\centerline{\includegraphics[width=17cm]{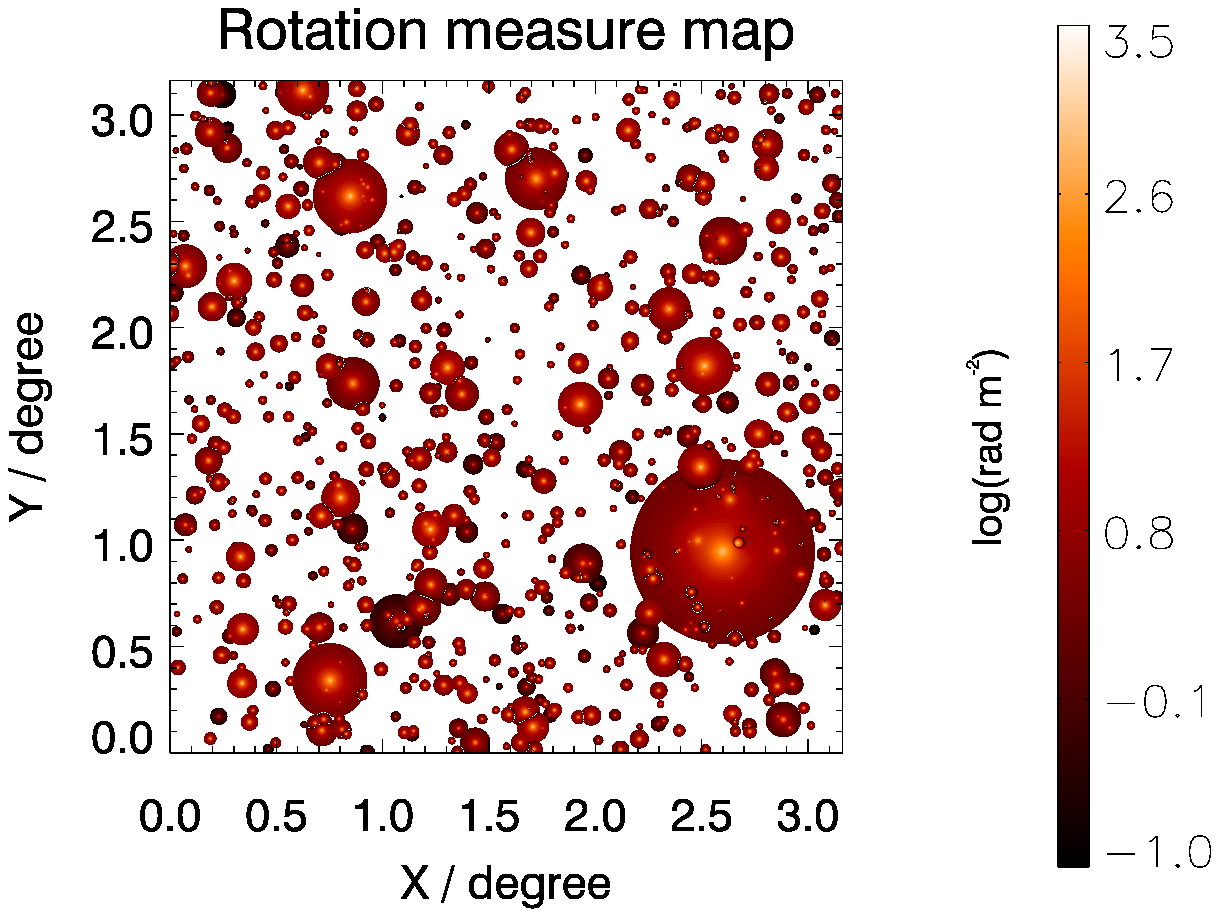}}
\centerline{\includegraphics[width=17cm]{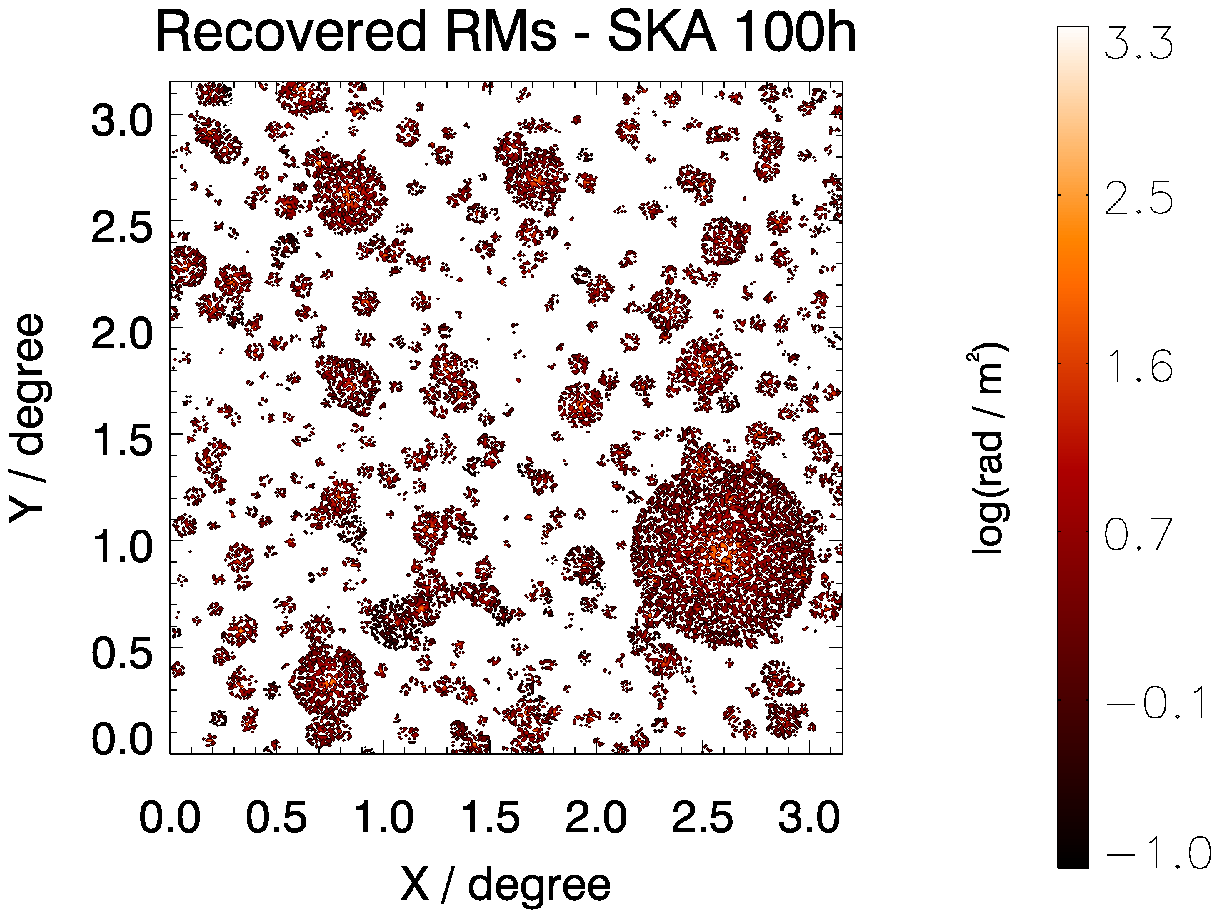}}
\caption{Top: Derived distribution of cosmological rotation measures on the sky
due to galaxy clusters. Bottom: Recovered RM-sky due to background radio
sources for a 100~h SKA-AA observation. 
\change{To enhance contrast, the resolution has
been reduced by a factor of ten compared to the figure on the top.} The real
field of view as planned to date would be 25 times larger.}\label{rm_dist}
\end{figure*}

\section{The simulations}\label{sims}

In order to assess the effect of cluster Faraday screens on the polarisation of
background radio sources, a relatively sophisticated model of clusters and
source populations is required.

\subsection{Cluster model}

In order to model the cluster number within an observed patch of sky (here 10
deg$^2$), we require a model for the cosmologically evolving cluster mass
function. For this, we employ a cluster mass function whose multiplicity
function \citep[see e.g.][]{ST99} is based on a moving barrier shape ansatz and
depends on a minimal number of parameters to provide a suitable fit to
numerical simulations. In particular, the shape of the used mass function has
been derived from N-body simulations of large cosmological volumes and high
resolutions following the method described in Jenkins et al.~(2001). The above
model gives the cluster mass function as a function of redshift. We adopt a
lower mass cutoff of $3\times 10^{13} \, M_\odot$ and for a given realisation
of the sky model derive cluster counts as a function of redshift, making
allowance for cosmic variance. Our resulting galaxy cluster catalogue contains
775 objects, with redshifts up to $\sim 2.3$. The properties of our generated
galaxy cluster sample are illustrated in Fig.~\ref{cls_props}.

In our model, clusters are distributed randomly and radio sources are either
background or foreground objects with respect to a given cluster. Every object
is placed on a 3D grid, covering $\sqrt{10} \times \sqrt{10}$ square degree and
the redshift space from zero to eight. The spatial resolution is $6000 \times
6000$ cells, and redshift space is covered by 240 cells. This gives an
effective resolution of 1.9 arcsec and d$z=0.033$ in the third dimension.

\subsection{Rotation measure model}

The observational data on the structure of magnetic fields in clusters is
currently sparse. Based on \citet{CKB01}, we therefore employ a simple model
for the rotation measure, with a peak value (justified in Section~\ref{coreRM}
below) drawn from a uniform random distribution between $RM_0 = \pm
5000$~rad~m$^{-2}$, and a decline according to a $\change{\beta}$-model:
\eq{\label{rmprof} \mathrm{RM}(r) =
   \mathrm{RM}_0 (1+r^2/r_0^2)^{-3\beta_\mathrm{\change{RM}}/2}.}
We use $\beta_\mathrm{\change{RM}}=1$ (see below) and $r_0=0.0636\, r_{200}$
\citep{BS2006}, where $r_{200}$ is the radius within which there is an
over-density of a factor 200 compared to the mean density of the universe at
the redshift of the cluster. $r_{200}$ is roughly equal to the virial radius of
the cluster. This results in core radii between 20 and 110~kpc, which
corresponds to angular sizes between a few arcseconds to half an arcminute.

\change{
\cite{BS2006} give a value of $\beta=2/3$ for the gas density distribution in
clusters of galaxies. The temperature distribution in clusters of galaxies is
typically not far from isothermal (except in the centres of cool core
clusters). Therefore, the pressure follows a very similar radial profile as that of
the density. Equipartition between the magnetic and the thermal pressure is
probably
a reasonable assumption for the magnetic field distribution
in clusters, and consistent with some RM-data \citep[e.g.][]{Govea06, Guiea08}.
Consequently, we adopt a value of $\beta_\mathrm{\change{m}}=1/3$ for the radial
distribution of the magnetic field strength along the line-of-sight. The
line-of-sight integral then results in the $\change{\beta}$-model for the 
rotation measure described in equation~(\ref{rmprof}) above, with 
$\beta_\mathrm{\change{RM}}=1$.}
We calculate the rotation measure contribution out to the virial radius. 

\begin{figure}
\centerline{\includegraphics[width=8.0cm]{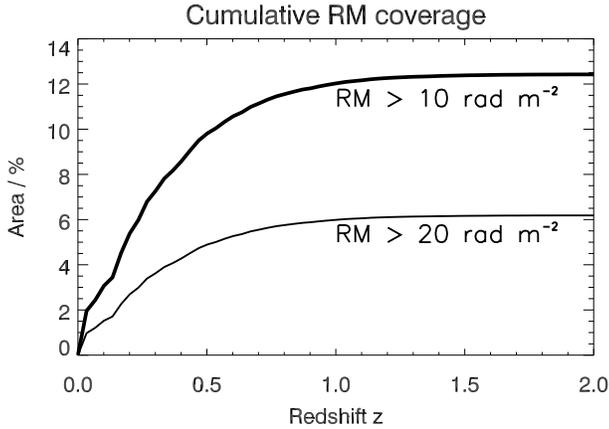}}
\caption{Fraction of sky covered with RM$>10$~rad~m$^{-2}$ (upper curve) and
RM$>20$~rad~m$^{-2}$ (lower curve).}\label{rm_cov}
\end{figure}

We show in Fig.~\ref{rm_dist} (top) the resulting distribution of rotation
measures due to the simulated galaxy clusters. Much of the sky is covered with
galaxy clusters, however, many of the systems have small rotation measures.
About 12~per~cent of the sky is covered with regions of at least
10~rad~m$^{-2}$. The fraction reduces by a factor of about two if one considers
rotation measures above 20~rad~m$^{-2}$. In our chosen cosmology, only
34~\change{per~cent} 
of all clusters are located beyond a redshift of one. These are
obviously smaller on average than the lower redshift population. Hence,
essentially all of the area coverage happens up to a redshift of unity
(Fig.~\ref{rm_cov}). Therefore, up to this redshift, structure determination
should be much easier than beyond.

\subsection{Radio source model}

To obtain realistic models of the radio source populations, we use the
publically available SKADS (Square Kilometre Array European Design Study)
synthetic radio source catalogue ({\em SKADS simulated skies, S$^3$};
\citealt{Wilmea08}\footnote{See: {\tt http://s-cubed.physics.ox.ac.uk/}.}). The
catalogue is derived from semi-empirical simulations of different populations
of radio continuum sources and extends down to faint flux limits in order to
allow observation simulations of high-sensitive future radio telescopes, such
as the SKA. The source populations included in the catalogue are radio-loud
({\sc FR\,i} and {\sc FR\,ii}) and radio-quiet AGNs as well as radio emitting
starburst and normal disk galaxies. The luminosity function corresponding to
each class of sources is derived empirically from available observations. The
fits are extrapolated to low luminosities and assume redshift evolutions to
assure the simulations are complete down to the instrumentally expected (faint)
flux limits. Our version of the catalogue covers ten square degrees of sky out
to a redshift of eight with a 1.4~GHz flux limit of 1~$\mu$Jy in Stokes~$I$. 
\change{See} \citet{Wilmea08} for a detailed description of
the radio source simulations.


\subsection{Polarisation model}

The degree of polarisation has to be determined for individual sources. We base
our prediction of the polarisation of these radio sources on the statistics of
the NRAO VLA Sky Survey (NVSS) catalogue \citep{Condea1998}. \citet{Tucea03}
have analysed polarisation statistics from the NVSS and other sources in
detail. Depending on spectral shape and source flux, they derive median
fractional polarisations between 1.12 and 1.77~per~cent for NVSS sources. For
the NVSS (1.4~GHz), the probability distribution of the fractional polarisation
declines monotonically, but for higher observing frequencies, a turnover may be
seen towards lower fractional polarisation. This is due to a systematic
increase of the fractional polarisation with observing frequency.
\citet{Tucea03} explain these findings by Faraday depolarisation. Therefore, we
expect that the low frequency distribution function also turns over at a not
yet measured fractional polarisation. We therefore fit a lognormal function to
the measured part of the fractional polarisation distribution function at
1.4~GHz. This results in a median value of 1.64~per~cent and a standard
deviation of 0.7~per~cent. \citet{Tucea03} believe that some of the NVSS
sources are depolarised by differential Faraday rotation in intervening galaxy
clusters. We argue that this would be much less so for SKA observations: The
main reason is that with lower flux limit the radio source population changes.
Extended jet sources dominate at high flux densities, and star forming galaxies
at increasing distance and decreasing angular size at lower flux densities. For
high redshift galaxy clusters, the depolarisation frequency is reduced by a
factor $(1+z)^{-1}$. Extended sources behind nearby clusters will be much
better resolved than by the NVSS (45~arcsec). For a turbulent power spectrum,
the RM variations should have less power on smaller scales, hence the
depolarisation frequency will be lower \citep{KAB2007}.

\change{For these reasons, we} 
adopt the de-depolarisation model from \citet{Tucea03}\change{
to derive the intrinsic polarisation $\Pi$ from the potentially Faraday 
depolarised polarisation $\Pi_\mathrm{FD}$, that is drawn from a random distribution 
as described above:}
\eqs{\Pi &=& \Pi_\mathrm{FD} f(\Pi_\mathrm{FD},\nu_\mathrm{GHz})
             / f(\Pi_\mathrm{FD},1.4) \\
f(p,\nu) &=& 72 \log(0.5 \nu^{0.4}+0.75) \exp(-3.2 p^{0.35}+0.8)\, .\nonumber}
\change{Here, $\nu_\mathrm{GHz}$ is a reference frequency in GHz.} 
We use 10~GHz
as the reference frequency, assuming that Faraday depolarisation for NVSS
sources mainly occurs below 10~GHz. The correction increases the fractional
polarisation by a factor of about two, for those sources with low polarisation
at 1.4~GHz, as shown in Fig.~\ref{fpol}. This effect has been measured already:
\citet{Taylea07} measure an increase of the fractional polarisation downwards
of 100~mJy of a factor of three compared to the high flux density sources, in
agreement with the model we propose.

\subsection{Redshift distribution}

\begin{figure}
\centerline{\includegraphics[width=8.0cm]{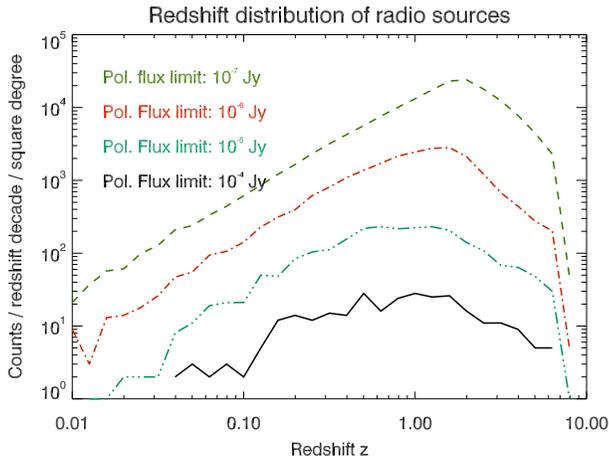}}
\caption{Distribution of radio sources against redshift for different polarised
flux limits. From top to bottom, the curves represent: 0.1, 1, 10, and
100~$\mu$Jy. The median redshift is 0.6, 0.6, 0.8, and 1.3.}\label{flsamps}
\end{figure}

For decreasing flux limit, the radio source counts are dominated by sources of
increasing redshift. We show the redshift distribution for different
(polarised) flux limited samples in Fig.~\ref{flsamps}. For a polarised
flux-density of 100~$\mu$Jy, roughly the level of current studies
\citep{Taylea07}, the median redshift is 0.6. It rises to 1.3 for a polarised
flux limit of 0.1~$\mu$Jy, appropriate for a 100~h SKA observation. Even for
the highest redshifts, we still predict thousands of background sources per
square degree for this flux limit.

\subsection{Required signal-to-noise for RM measurements}\label{snr_req}

A distinctive feature of the next generation of radio telescopes is that a
large bandwidth and high spectral resolution will be available simultaneously.
Rotation measure synthesis is an established technique to analyse RM data.
However, for the analysis of the RM-grid, we wish to extract information from
the faintest detectable sources. For the faintest sources we will extract a
single value of the rotation measure. Here, we consider the minimum
signal-to-noise on a point source required for this. Faraday rotation leads to
sinusoidal variation with respect to $\lambda^2$ across the band for the Stokes
parameters $Q$ and $U$, modulated by a function giving the fractional
polarisation as a function of wavelength and the intrinsic spectrum of the
source. We consider normalised Stokes parameters:
\eqs{\left(\begin{array}{r}Q^\prime \\ U^\prime\end{array}\right)
    &=&(Q^2+U^2)^{-1/2} 	\left(\begin{array}{r}Q \\ U\end{array}\right)\label{qu}\\
        &=&\left(\begin{array}{r}\cos(2\mathrm{PA}+2\mathrm{RM} \, \lambda^2)\\
      \sin(2\mathrm{PA}+2\mathrm{RM} \, \lambda^2)\end{array}\right),\nonumber}
where $Q$ and $U$ are the ordinary Stokes parameters. To assess the required
signal-to-noise ratio (SNR), required of a background source to measure a given
foreground RM, we perform Monte-Carlo simulations. Our simulated data were for
a background source with a spectral index of~0.7 -- we simulate the observing
band, taking a constant channel width of $\Delta\nu = 1$~MHz. We perform a
combined $\chi^2$-fit to the normalised Stokes parameters, and use the flux at
1.4~GHz as a reference point for the SNR. For infinite SNR, the minimisation of
$\chi^2$ would be straight forward. For decreasing SNR, additional minima
appear in $\chi^2$ as a function of RM. For a SNR of 10 (100) the distance
between these minima is of order 1~rad~m$^{-2}$ (100~rad~m$^{-2}$). A dense
grid of starting guesses is therefore essential, for a gradient search method
to find the global minimum. The actual values of $\chi^2$ differ only by a few
percent, with the exception of the $\chi^2$ for the true RM, and its negative.
We find that $\chi^2$ for the true RM and its negative, are significantly lower
than all other RMs (five times the standard deviation for 400 starting guesses)
for a signal to noise of four and higher.

\change{
The sign ambiguity is due to the symmetry properties of the trigonometric
functions. If a set $(\mathrm{PA,RM})$, where PA is the high frequency
polarisation angle (compare equation (\ref{qu})), are the correct parameters to
describe both functions, $Q^\prime(\mathrm{PA,RM})$ and
$U^\prime(\mathrm{PA,RM})$, then $(\mathrm{\pi+PA, -RM})$ will be another
solution regarding $Q^\prime$ and $(\mathrm{-PA, -RM})$ respectively for
$U^\prime$:
\eqs{Q^\prime(\mathrm{PA,RM})&=&Q^\prime(\mathrm{\pi+PA,-RM})\nonumber\\
  \nonumber U^\prime(\mathrm{PA,RM})&=&U^\prime(\mathrm{-PA,-RM})}
To fix the sign at low SNR, for each set of fitted parameters for $Q^\prime$ we
also performed two fits for $U^\prime$ with starting guesses of
$(\mathrm{\pi-PA, RM})$ and $(\mathrm{PA, RM})$, respectively, allowing only RM
to vary, i.e.\ keeping the high frequency polarisation angle fixed. If the fit
with $(\mathrm{PA, RM})$ leads to the lowest minimum of $\chi^2$, then we
have found the correct RM. However, if the fit with $(\mathrm{\pi-PA, RM})$
produces the lowest minimum of $\chi^2$, then $-$RM will be the correct
rotation measure. The procedure is then repeated with $U^\prime$ and $Q^\prime$
exchanged. We may thus produce a combined $\chi^2$ for RM and $-$RM.} Selecting
on the combined $\chi^2$, we are able to recover the correct sign.

For RMs between 3 and 3000~rad~m$^{-2}$, we find a minimum SNR between four and
ten. This is true for the observing bands from 800~MHz to 1050~MHz as well as
300~MHz to 1000~MHz. In the following, we assume that a RM measurement may be
done for any source detected at an SNR of ten in polarised flux.

\subsection{Rotation measure grid}

\begin{figure}
\centerline{\includegraphics[width=8.0cm]{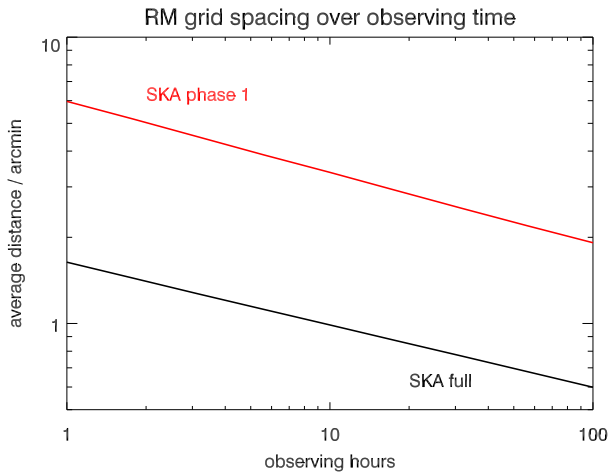}}
\caption{RM-grid spacing for different telescopes and a SNR of
ten.}\label{spacing}
\end{figure}

For observations of nearby clusters, as well as the galaxy, it is useful to
compute the root-area-average distance between polarised background sources.
Fig.~\ref{spacing} shows the average grid spacing over observing time for the
adopted SKA and SKA phase~1 sensitivities. We use the point source sensitivity
limit according to \citet{WW99}:
\eq{\mathrm{d}s=\frac{k_\mathrm{B} T_\mathrm{sys}}{\sqrt{2 \,\mathrm{d}\nu\,
      \tau}\,A\, \epsilon}\, ,}
where, $k_\mathrm{B}$ is Boltzmann's constant, $T_\mathrm{sys}$ is the
telescope's system temperature, $A$ the total effective area, $\mathrm{d}\nu$
the bandwidth,  $\tau$ the observing time and $\epsilon$ the overall
efficiency. Adopting a SNR of ten, we count the sources above the sensitivity
limit for a particular telescope and observing time. The total source counts
are roughly inversely proportional to the polarised flux, which is an effect of
the adding up of the contributions from different source types. So, while the
galaxies constitute the majority of the sources at low fluxes, other source
types still contribute enough to flatten the overall distribution considerably.

\subsection{Expected RM values}

\begin{figure}
\centerline{\includegraphics[width=8.0cm]{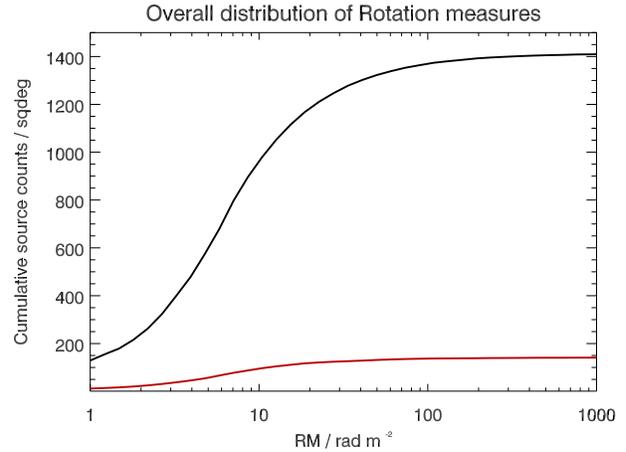}}
\caption{Cumulative histogram of all the rotation measures we expect to find
within one square degree for a polarised flux limit of 1~$\mu$Jy (lower red
curve) and 0.1~$\mu$Jy (upper black curve). In any case, we expect most of the
measurements to yield rotation measures of order 10 to one hundred
rad~m$^{-2}$.}\label{rmhisto}
\end{figure}

For each background source, we calculate the observed rotation measure by
integrating through the simulation volume from the observed source to the
observer. We show the cumulative histogram of all the rotation measures
produced in this way in Fig.~\ref{rmhisto}. We find that the majority of
rotation measures are in the range 10--100~rad~m$^{-2}$, due to the fact that
most of the area comes from the outskirts of the clusters. The highest rotation
measures come of course from the cluster centres. For a given frequency band
and rotation measure range, a particular SNR is required. As shown above in
Section~\ref{snr_req}, both of the frequency ranges, 300~MHz to 1000~MHz
(aperture array) and 800~MHz to 1050~MHz (dish array), should be suitable for
the great majority of expected rotation measures.

\subsection{Core rotation measures}\label{coreRM}

\begin{figure}
\center{\includegraphics[width=8.0cm]{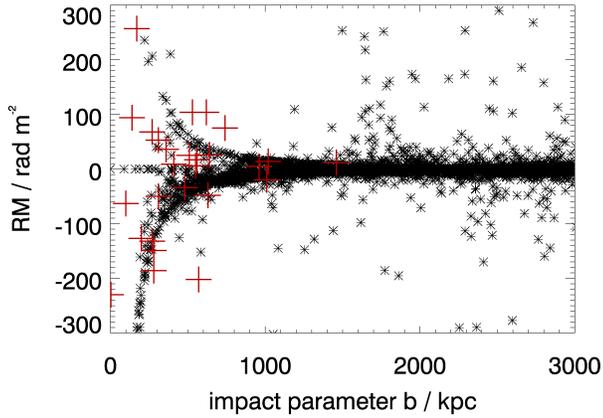}}
\caption{Rotation measure against impact parameter. The impact parameter is the
\change{projected} 
distance of a radio source to a given cluster centre. Low redshift
($z<0.5$), high mass ($M>10^{14}\,M_\odot$) clusters have been selected in
order to compare to the observed data of \citet{CKB01}. The polarised flux
limit is 1~$\mu$Jy, as in \citet{CKB01}. Observational data from \citet{CKB01}
(cluster sample only) is provided as red crosses. For distances greater than
1000~kpc, the sightlines to the radio sources pierce neighbouring clusters. To
get the observed RM spread at the observed impact parameters, central RMs of
about 5000~rad~m$^{-2}$ are necessary.}\label{RMR}
\end{figure}

For each radio source within three virial radii of a given cluster centre, we
also calculate the respective impact parameter, i.e.\ the projected distance to
the cluster centre. We show these rotation measures against the impact
parameter, for low redshift ($z<0.5$), high mass ($M>10^{14}\,M_\odot$)
clusters and a polarised flux limit of 1~$\mu$Jy in Fig.~\ref{RMR}. These cuts
were employed in \citet{CKB01}, and their data is also shown in Fig.~\ref{RMR}.
\change{But our model and the observational data show only RM values up to
$|\mathrm{RM}| < 300$~rad~m$^{-2}$, whereas our model has peak values of
5000~rad~m$^{-2}$ in the cluster centres. The reason is that at the high flux
cut adopted here, the RM-grid is still very sparse, and hardly any source is
located within the core radius. If we employ a lower central RM, with the
same cuts, the simulated RM spread on the scale of the virial radius would be
too small compared to the observed data. As can be seen, the central rotation
measures of about 5000~rad~m$^{-2}$ that are used in the simulated data here
are necessary to reproduce the observed spread of rotation measures.} For
distances greater than 1000~kpc and random sightlines, we encounter high RMs
from neighbouring clusters. These are not present in \citet{CKB01}, as they
select their control sample avoiding known X-ray clusters.

\section{Results}\label{res}


\subsection{Detection statistics}\label{stat}

\begin{figure*}
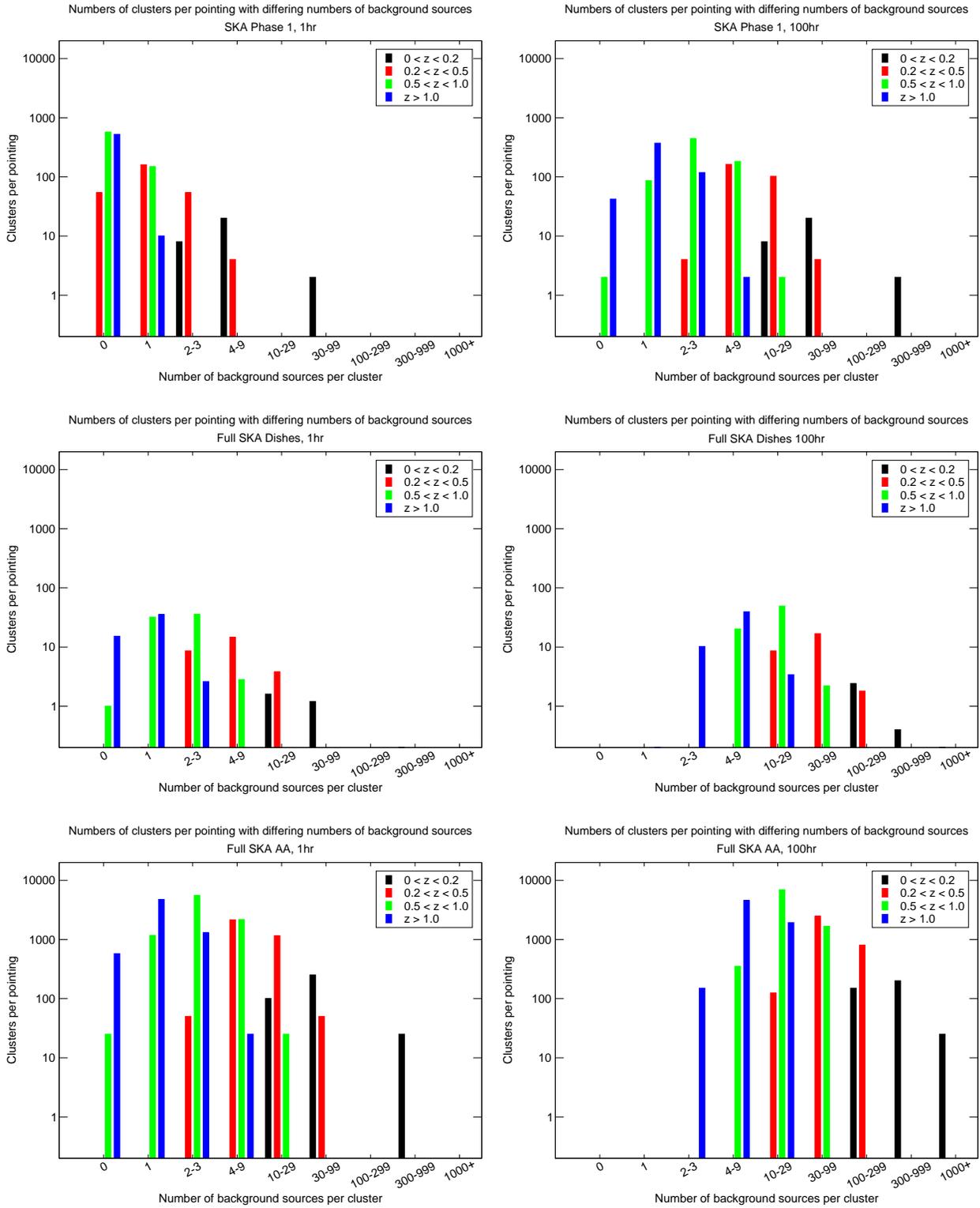

\centerline{\includegraphics[angle=270,width=8.0cm]{SKA_P1_1hr.epsi}
      \quad \includegraphics[angle=270,width=8.0cm]{SKA_P1_100hr.epsi}}
\vspace*{5mm}
\centerline{\includegraphics[angle=270,width=8.0cm]{SKA_Dish_1hr.epsi}
      \quad \includegraphics[angle=270,width=8.0cm]{SKA_Dish_100hr.epsi}}
\vspace*{5mm}
\centerline{\includegraphics[angle=270,width=8.0cm]{SKA_AA_1hr.epsi}
      \quad \includegraphics[angle=270,width=8.0cm]{SKA_AA_100hr.epsi}}
\caption{Detection statistics for the SKA aperture array (bottom row), the full
SKA dish array (middle row) and the phase~1 SKA dish array (top row), telescope
specs from Table~\ref{telcomp}, for an observing time of 1~hour (left) and
100~hours (right), respectively. We plot the number of background sources per
cluster on the horizontal axis, and the cluster counts for the particular bins
on the vertical axis. The huge number counts found with the full SKA aperture
array is a merit of its huge field of view (250 square degrees) combined with
its high sensitivity. The same information is provided in tabular form in
Table~\ref{dstatt}.}\label{dstat}
\end{figure*}

We show the detection statistics for the SKA aperture array and the phase~1~SKA
in Fig.~\ref{dstat} and Table~\ref{dstatt}. The histograms demonstrate the
progress that can be expected from the SKA, especially the full aperture array.

For a one hour phase~1 dish blind survey, we predict to find 30, 214, and 148
clusters in the redshift ranges 0--0.2, 0.2--0.5, 0.5--1, respectively, with
\change{more than one background RM measurement}, each. 
We also predict a few detections
at higher redshift. This would allow stacking experiments for different
redshift bins out to a redshift of at least one. Also, one could determine the
decline of the average RM with cluster radius better than presently possible.
This could answer the question weather the magnetic field energy density
follows the thermal energy density. We also would have some chance to find
about a couple of nearby clusters with 30--100 RMs, allowing an improved
determination of the field structure. 30 background sources should be enough to
constrain the power spectrum \citep{Govea06,Guiea08}. Since the positions of
the big nearby clusters are known, a targeted survey would certainly yield
about this number of RMs.

A 100~hour phase~1 observation would result in 26 clusters \change{at} 
$\change{z<0.5}$ with more
than 30 background sources, and a couple of these having more than 300
background sources. This would allow to study the field structure in some
detail. Up to a redshift of unity, nearly every cluster in the field of view
should have at least one background source. Even beyond $z=1$, we predict to
find 490~clusters with one \change{to several} background sources. This would allow the
extension of the stacking method to this redshift.

The full SKA aperture array will finally allow to measure the structure of the
magnetic field in great detail: Even for a 1~hour observation, each low
redshift cluster will at least have 10 background sources 25 of which over
\change{three} hundred. 
We predict to find \change{325} clusters in total with more than 30
background sources, distributed over a redshift range up to~0.5, promising good
statistics of the variance of the power spectrum from cluster to cluster. More
than 10,000 clusters would be detected with at least one RM measurement beyond
a redshift of 0.5.

\begin{figure}
\centerline{\includegraphics[width=8.0cm]{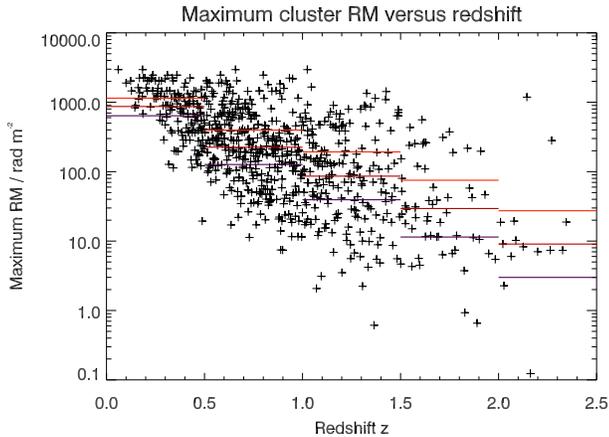}}
\caption{Maximum rotation measure seen against any given cluster in our sample
against cluster redshift for a polarised flux limit of 0.1~$\mu$Jy. The red
lines denote the median for sub-samples with corresponding redshifts (0--0.5,
0.5--1, 1--1.5, 1.5--2, $>2$). The purple lines denote the measured rotation
measures if the true rotation measures would decline with redshift as
$(1+z)^{-1}$, the orange line displays the medians for an increase proportional
to $(1+z)$.}\label{rmmaxz}
\end{figure}

For a 100~hour pointing, we will be able to constrain the power spectrum for
many clusters in the field of view up to a redshift of one ($>5,000$ clusters
with more than 30 background sources). Up to a redshift of one, we expect
$\change{>1,000}$ 
clusters with at least 100~RMs. Only for such clusters, we may expect
to {\em resolve} the core in some cases, i.e.\ to have at least one background
source within the cluster's core radius. A 100~hour pointing would correspond
roughly to the sensitivity limit used in Fig.~\ref{rmmaxz}, where we show the
decrease of the maximum rotation measure in any given cluster with redshift due
to the decreased likelihood to get a background source within the core radius.
25 clusters should have more than 1000 background sources, and hence allow a
very detailed determination of the magnetic field structure, even in the
cluster cores.

Due to the limited field of view, the full SKA dish array would not nearly
reach the detection rates of the aperture array. Even a 100~hour blind survey
would only yield \change{24} clusters with more than 30 background sources.
%

\begin{table*}
\begin{minipage}{12.5cm}
\caption{Cluster counts with a given number of background
sources.}\label{dstatt}
\begin{tabular}{lrrrrrrrrr}\hline
	Redshift & \multicolumn{9}{c}{Number of polarised background sources per cluster} \\		
        interval & 0 	& 1 	& 2--3	& 4--9  & 10--29  & 30--100  &  100--299  & 300--999  &  $1000+$ \\\hline
\multicolumn{10}{c}{Phase 1 Dishes, 1 hour pointing} \\\hline
        0--0.2   &0	&0	&8	&20	&0	&2	&0	&0	&0      \\
        0.2--0.5 &54	&160	&54	&0	&0	&0	&0	&0	&0	\\
        0.5--1   &568	&148	&0	&0	&0	&0	&0	&0	&0	\\
        $>1$    &522	&10	&0	&0	&0	&0	&0	&0	&0	\\\hline
\multicolumn{10}{c}{Phase 1 Dishes, 100 hour pointing} \\\hline
        0--0.2   &0	&0	&0	&0	&8	&20	&0	&2	&0	\\
        0.2--0.5 &0	&0	&4	&162	&102	&4	&0	&0	&0	\\
        0.5--1   &2	&86	&444	&182	&2	&0	&0	&0	&0	\\
         $>1$   &42	&370	&118	&2	&0	&0	&0	&0	&0	\\\hline
\multicolumn{10}{c}{Full SKA dishes, 1 hour pointing}\\\hline
        0--0.2   &0	&0	&0	&0	&1.6	&1.2	&0	&0.2	&0	\\
        0.2--0.5 &0	&0.2	&8.6	&14.6	&3.8	&0	&0	&0	&0	\\
        0.5--1   &1	&32	&35.8	&2.8	&0	&0	&0	&0	&0	\\
         $>1$   &15.2	&35.4	&2.6	&0	&0	&0	&0	&0	&0	\\\hline
\multicolumn{10}{c}{Full SKA dishes, 100 hour pointing}\\\hline
        0--0.2   &0	&0	&0	&0	&0	&0	&2.4	&0.4	&0.2	\\
        0.2--0.5 &0	&0	&0	&0	&8.6	&16.8	&1.8	&0	&0	\\
        0.5--1   &0	&0	&0	&20.2	&49.2	&2.2	&0	&0	&0	\\
         $>1$   & 0	&0.2	&10.2	&39.4	&3.4	&0	&0	&0	&0	\\\hline
\multicolumn{10}{c}{Full SKA AA, 1 hour pointing} \\\hline
        0--0.2   &0	&0	&0	&0	&100	&250	&0	&25	&0	\\
        0.2--0.5 &0	&0	&50	&2150	&1150	&50	&0	&0	&0	\\
        0.5--1   &25	&1175	&5550	&2175	&25	&0	&0	&0	&0	\\
         $>1$   &575	&4750	&1300	&25	&0	&0	&0	&0	&0	\\\hline
\multicolumn{10}{c}{Full SKA AA, 100 hour pointing} \\\hline
        0--0.2   &0	&0	&0	&0	&0	&0	&150	&200	&25     \\
        0.2--0.5 &0	&0	&0	&0	&125	&2475	&800	&0	&0	\\
        0.5--1   &0	&0	&0	&350	&6925	&1675	&0	&0	&0	\\
         $>1$   & 0	&0	&150	&4575	&1925	&0	&0	&0	&0	\\\hline
\end{tabular}
\end{minipage}
\end{table*}

\subsection{Cosmological evolution}

Clusters at higher redshift cover less area of sky (up to a redshift of 1.6).
Also, there are fewer sources behind a unit area at higher redshift. Therefore,
the probability that a cluster centre has a background object decreases as the
redshift increases. We plot the maximum rotation measure that we detect for a
polarised flux limit of 0.1~$\mu$Jy against any given cluster versus the
cluster redshift in Fig.~\ref{rmmaxz}. Out to a redshift of unity, there is
still a good chance of finding sightlines through the cluster core. However,
the median maximum rotation measure decreases by almost a factor of four for
clusters with redshift 0.5--1 compared to ones at 0--0.5. The decrease in the
median of the measured maximum rotation measure is almost exponential with
redshift, with an $1/$e-decay length in $z$ close to 0.4. We can check the
detectability of possible trends with redshift by multiplying the rotation
measures with $(1+z)^n$ where we have chosen n to be $\pm1$ as example. The
change in the median rotation measure is also indicated in Fig.~\ref{rmmaxz}.
The decay length changes to 0.3 and 0.5, respectively. If we compare the
redshift intervals 0--0.5 and 0.5--1, we predict a decrease of the median
rotation measure by factors of about 3,4 and 5, respectively (compare with
Table~\ref{rmzcomp}, which is normalised to the low redshift value).

The statistical error on the median maximum rotation measure is given by:
\dm{\sigma_\mathrm{M}= 1.253 \sigma/\sqrt{N},}
where $\sigma$ is the standard deviation and $N$ is the number of sources in
the respective bin. We have calculated the median maximum RM for any given
cluster and the statistical errors for a 100~h SKA~AA observation for different
values of $n$. This is shown in Table~\ref{rmzcomp}. Differences in the
exponent $n$ of $0.3$ could be distinguished.

\section{Discussion}\label{disc}

We assess the capability of the SKA and its precursor, phase~1 SKA, to detect
rotation measures and determine the evolution of magnetic fields in galaxy
clusters. Our cluster and radio source simulations are based on standard
cosmological methods and extrapolation of available observational data. We use
a fractional polarisation model based on NVSS data, moderately corrected for
Faraday depolarisation, at low fractional polarisations, only.

We find that about 10~per~cent of the whole sky should be covered with rotation
measures greater than 10~rad~m$^{-2}$. If the cosmology is known, this may
serve as an additional constraint to check for the proper removal of the Milky
Way's RM foreground. Large clusters can have a RM contribution from background
clusters. However, the contamination occurs mainly on the outskirts of these
clusters.

Our simulations show that core RMs of about 5000~rad~m$^{-2}$ are necessary to
explain the currently known stacking data, which shows about 200~rad~m$^{-2}$
on the scale of the virial radius. RMs against central cluster radio sources
often reach values comparable to our core RMs \citep{CT02}. This suggests that
our assumption for the core RMs is realistic, but also that the functional form
of the radial decline we assume is not too far from reality.

\change{
How would the derived statistics change if one were to change the radial
RM-profile? The adopted profile has high core values with a steep decline.
With this distribution we have shown to match the observations of nearby clusters.
However, one might argue that these nearby clusters may not be representative 
of whole cluster population. One might imagine that a less centrally peaked 
flatter distribution, like for example in the Coma cluster, might be more typical.
We have shown in section~\ref{snr_req} above, that the optimum RM-range is between a few and a several 
100~rad~m$^{-2}$. Such a scenario would therefore lead to an increase of the 
fraction of sky in our best range. Therefore,  the RM-distribution
we have chosen is, if anything, rather pessimistic.}

\begin{table*}
\begin{minipage}{10.5cm}
\caption{Median maximum rotation measure in any given cluster against redshift
for different assumptions about the intrinsic evolution, where $n$ is the
exponent for a systematic change with redshift. The statistical errors are
applicable for a 100~hour SKA~AA observation. \change{The values in each row have been
normalised to the lowest redshift bin.}}\label{rmzcomp}
\begin{tabular}{rrrrrr@{}}\hline
  $n$ & $z=0{-}0.5$  & $z=0.5{-}1$     & $z=1{-}1.5$     & $z=1.5{-}2$     & $z>2$           \\ \hline
   1  & $1\pm0.015$  & $0.346\pm0.010$ & $0.168\pm0.012$ & $0.066\pm0.009$ & $0.024\pm0.058$ \\
  0.5 & $1\pm0.015$  & $0.298\pm0.009$ & $0.128\pm0.009$ & $0.047\pm0.007$ & $0.016\pm0.038$ \\
  0.2 & $1\pm0.015$  & $0.276\pm0.008$ & $0.109\pm0.007$ & $0.039\pm0.005$ & $0.012\pm0.029$ \\
   0  & $1\pm0.015$  & $0.259\pm0.008$ & $0.099\pm0.007$ & $0.034\pm0.005$ & $0.010\pm0.024$ \\ \hline
\end{tabular}
\end{minipage}
\end{table*}

We show that with a phase~1 SKA, as defined in Table~\ref{telcomp}, good
stacking experiments should be possible, determining the average magnetic field
profile locally and also out to a redshift of about one. The field structure
should be measurable for some nearby clusters. If there would be a choice
necessary between a deep survey and 100~one hour pointings, one should probably
do the deep survey, since it already allows some field structure determination
in nearby clusters, and promises to measure RMs for clusters with $z>1$.

With the full SKA~AA aperture array, both a shallow 100 one hour pointing
survey, and a deep survey promise very interesting results. A shallow 100~times
one hour survey should detect RMs from about $10^6$ clusters, with complete
redshift coverage. A deep 100~h one field survey will allow structure
determination in unprecedented detail out to a redshift of one: over 1000
clusters will have more than 100~RM measurements, 25 of which over 1000 RMs.
\change{
The result of a 100~h deep survey is also shown graphically in
Fig.~\ref{rm_dist} (bottom). To some extent, the two approaches are
complementary: the 100~h deep survey would offer a complete RM survey for all
structures above $3\times 10^{13} \, M_\odot$, and thereby exclude possible
selection biases which could arise if only a fraction of the clusters is
detected as in the shallow survey. The shallow survey would be able to detect
large scale bias, and also would provide better overall statistics on the
cosmological field evolution, if details within the clusters are neglected.}

Without phased array feeds on the majority of the dishes, as assumed here, the
detection rates remain rather small for the full SKA dish array. However,
having similar sensitivity than the aperture array, but coverage at higher
frequencies, it would be a valuable complement for high RM regions (cluster
centres), where the frequencies accessible to the aperture array would be
depolarised.The RM grid density would be similar to the aperture array.

With the deep survey, we could measure the cosmic evolution of the RMs in
clusters. If RM is proportional to $(1+z)^n$, we could measure $n$ to an
accuracy of ${\Delta}n=0.3$. In order to determine the evolution of the
magnetic field from this, one needs additional information on the electron
densities. If we would adopt an accuracy for $n$~of ${\Delta}n=0.3$, and assume
a similar behaviour and accuracy for the electron densities, to be measured
via the Sunyaev--Zel'dovich effect \citep{SKH05}, we would end up with an
accuracy for a power law index for the magnetic field ($B\propto(1+z)^m$) of
$0.4$. This should be regarded as a conservative estimate, since it only takes
into account information from the deep survey. Additional information from a
shallow survey should improve the accuracy for the low redshift bins. Also, we
have only used the information on the maximum RM per cluster. Taking into
account the full data, will also improve the statistics.

For a polarised flux limit of 0.1~$\mu$Jy, as appropriate for a 100 hour SKA
pointing and a SNR of ten, we predict a few 10,000 radio sources per square
degree. These sources have a broad redshift distribution, with a median
redshift of about one, rising with decreasing flux limit. We can hence expect
not only to find the reported numbers of background sources, but also a sizable
number of foreground sources, of order a few percent of the background sources,
rising of course with cluster redshift. An accurate determination of the
foreground RM opens up the possibility to detect a possible small mean cluster
RM which would point to a super-cluster scale field connecting the cosmic web.
This would be the first detection of such a field.

\section{Conclusions}\label{conc}

We have modelled the distribution of rotation measures for the SKA. We
calculate the average distance between rotation measures to be between
6~arcminutes for a one hour observation with the early SKA and 36~arcseconds
for a hundred hour pointing with the full SKA. We expect to find rotation
measures mostly up to 100~rad~m$^{-2}$, with the cluster centres reaching up to
several thousand. The planned SKA mid frequency aperture array (300--1000~MHz)
\change{
as well as the lowest band of the dish array} would be well suited for 
the great majority of sources. High RM cluster centres would
require targeted high frequency follow up observations. We find that current
Faraday rotation studies only represent the outskirts of galaxy clusters. The
average RM should increase steadily towards the core to reach typically several
thousand rad~m$^{-2}$.

A phase~1 SKA would already improve the statistics for current stacking
experiments considerably, and make this experiment viable out to redshifts
$z>1$. The full SKA aperture array would detect over a million clusters with at
least one background source each in a shallow 100~hour survey, and allow
detailed field structure determination ($>1000$ clusters with more than 100
background sources each) with a deep survey. If the cosmological evolution of
the rotation measures is proportional to $(1+z)^n$, the SKA would be able to
measure $n$ to an accuracy of 0.3. Compared to the few RMs known for a few
nearby clusters today, this will revolutionise our knowledge of rotation
measures in galaxy clusters. Provided the electron densities can be measured
via the Sunyaev--Zel'dovich effect at high redshift, we can expect to follow
the build-up of cosmic magnetism with the SKA.

\section*{Acknowledgments}
This activity was (partly) supported by the European Community Framework
Programme~6, Square Kilometre Array Design Studies (SKADS), 
contract no 011938.

\bsp
\setlength{\labelwidth}{0pt}

\label{lastpage}

\end{document}

%% file: def.tex
\bibpunct{(}{)}{;}{a}{}{,}

\newcommand{\ignore}[1]{}
\newcommand{\eq}[1]{\begin{equation} #1 \end{equation}}

\newcommand{\eqs}[1]{\begin{eqnarray} #1 \end{eqnarray}}

\newcommand{\change}[1]{{ #1}}

\newcommand{\dm}[1]{\begin{displaymath} #1 \end{displaymath}}